\documentclass[a4paper,
               keeplastbox,  
               ]{jacow}
\usepackage{pdfpages,multirow,ragged2e} %

\makeatletter
	\ifboolexpr{bool{xetex}}
	 {\renewcommand{\Gin@extensions}{.pdf,%
	                    .png,.jpg,.bmp,.pict,.tif,.psd,.mac,.sga,.tga,.gif,%
	                    .eps,.ps,%
	                    }}{}
\makeatother

\ifboolexpr{bool{xetex} or bool{luatex}} 
 {}                                      
 {\usepackage[utf8]{inputenc}}           

\usepackage[USenglish]{babel}

\ifboolexpr{bool{jacowbiblatex}}%
 {
  \addbibresource{jacow-test.bib}
  \addbibresource{biblatex-examples.bib}
 }{}
\usepackage{enumitem}
\usepackage{tikz}
\usepackage{pgfcalendar}
\usepackage{datetime2}
\usepackage{adjustbox}  
\usepackage{xcolor}
\definecolor{tanngruen}{RGB}{3, 150, 80}

\begin{document}

\title{FAIR Commissioning - Towards First Science}

\author{S. Reimann\thanks{s.reimann@gsi.de},
H. Albers, R. W. Assmann\textsuperscript{1}, P. Gasik, O. Geithner, F. Hagenbuck, A. Herlert,  \\
P. Hofmann, V. Kamerdzhiev, M. Kauschke, H. Kollmus, S. Pietri, N. Pyka, T. Radon, C. Schroeder, \\
S. Schwarz, H. Simon, P. Spiller, K. Vogt\\
GSI Helmholtz Centre, Darmstadt, Germany\\
\textsuperscript{1}also at Institute for Applied Physics (IAP), Goethe University Frankfurt, Frankfurt, Germany}
	
\maketitle

\begin{abstract}
The international Facility for Antiproton and Ion Research (FAIR) is under construction at the GSI Helmholtz Centre in Darmstadt. The first project stage includes the superconducting \SI{100}{\tesla\meter} heavy-ion synchrotron SIS100, the Super Fragment Separator, and associated beam transport lines. Part of GSI’s existing accelerator chain, comprising UNILAC and SIS18, will serve as injector. Installation work in the FAIR accelerator tunnels and supply buildings has been ongoing since early 2024. As progress continues, special attention is now on the start of commissioning, beginning in 2025 with the cryogenic plant. Commissioning of the transport line will follow at the end of 2025, and beam commissioning is scheduled for the second half of 2027. This paper outlines the current status of the project, commissioning strategy and timeline. 

\end{abstract}

\section{GSI \& FAIR}
\begin{figure}[!htb]
   \centering
   \includegraphics*[width=1.0\columnwidth]{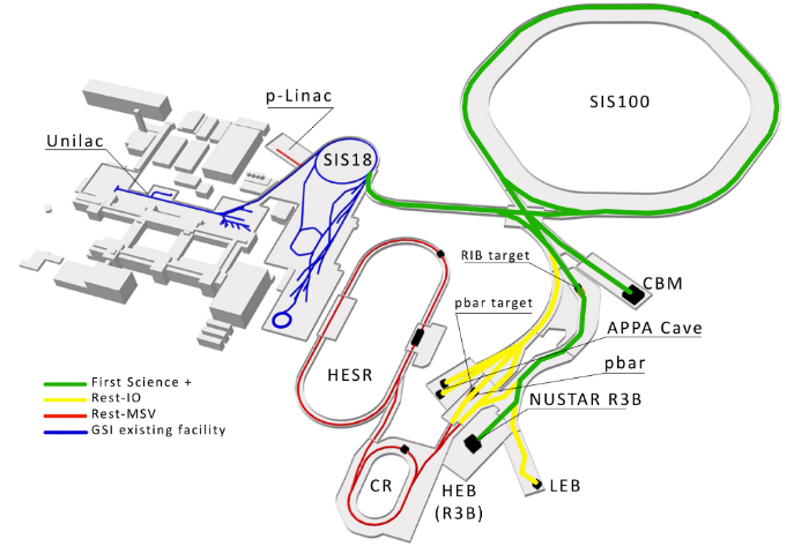}
   \caption{Overview of the GSI accelerator complex (in blue) and the facilities of the FAIR project. The first project stage, 'First Science +', is depicted in green.}
   \label{fig:gsifair}
\end{figure}

Figure~\ref{fig:gsifair} shows the existing GSI accelerator complex and the FAIR~\cite{Spiller:IPAC18-MOZGBF2, blaurock:ipac2023-thyd1} project facilities currently under construction. The first stage being implemented is referred to as “First Science+”. Within this stage, beam delivery from SIS18 via the SFRS to the high-energy branch of the NUSTAR cave will be the first to be commissioned; this marks the “Early Science” milestone. The FAIR Phase-0 user program~\cite{Bai2018} will continue to be carried out by GSI throughout the entire commissioning phase.
\section{The Cryogenic System}
\begin{figure}[!htb]
   \centering
   \includegraphics*[width=1.0\columnwidth]{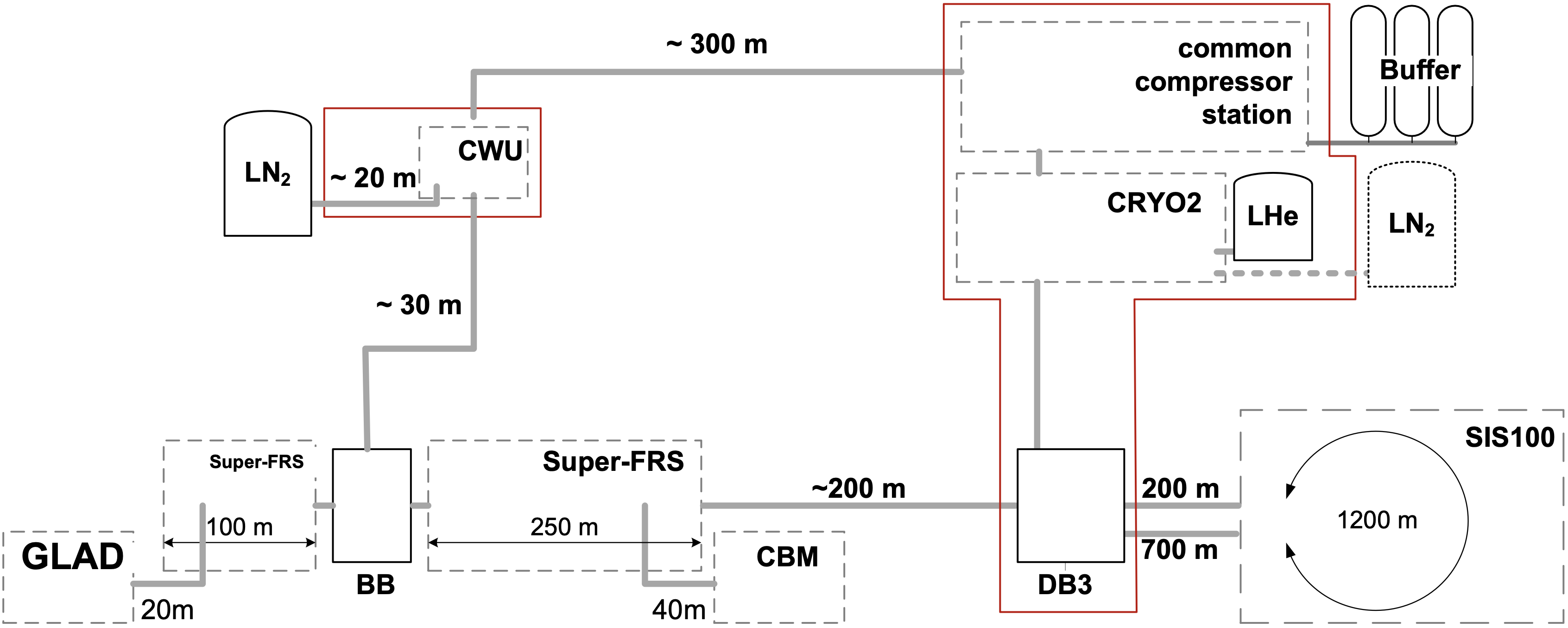}
   \caption{Layout of the FAIR cryogenic system with cryo plant and corresponding distribution system.}
   \label{fig:cryo}
\end{figure}

The FAIR cryogenic system~\cite{mcentee2023cryogenics} (Fig.~\ref{fig:cryo}) is designed to provide \SI{14}{\kilo\watt} of cooling power at \SI{4.4}{\kelvin} and \SI{49}{\kilo\watt} at \SI{50}{\kelvin}, supplying two major machines: the fast-ramping heavy-ion synchrotron SIS100 and the Super Fragment Separator, as well as the superconducting magnets~\cite{Fischer:IPAC17-WEOCB2} of CBM and NUSTAR. Commissioning of the cryoplant "Cryo2" is scheduled to begin in June 2025, initially in connection with the first distribution box (DB3). This phase involves two buildings, the cryoplant and the compressor station, along with helium storage tanks holding an initial helium inventory of approximately \SI{10000}{\cubic\meter}. Due to ongoing construction in the surrounding area, special safety measures will be implemented. Subsequently, the distribution systems for the Super-FRS and SIS100 will be commissioned.

\section{FAIR Control Centre (FCC)}
\begin{figure}[!htb]
   \centering
   \includegraphics*[width=0.91\columnwidth]{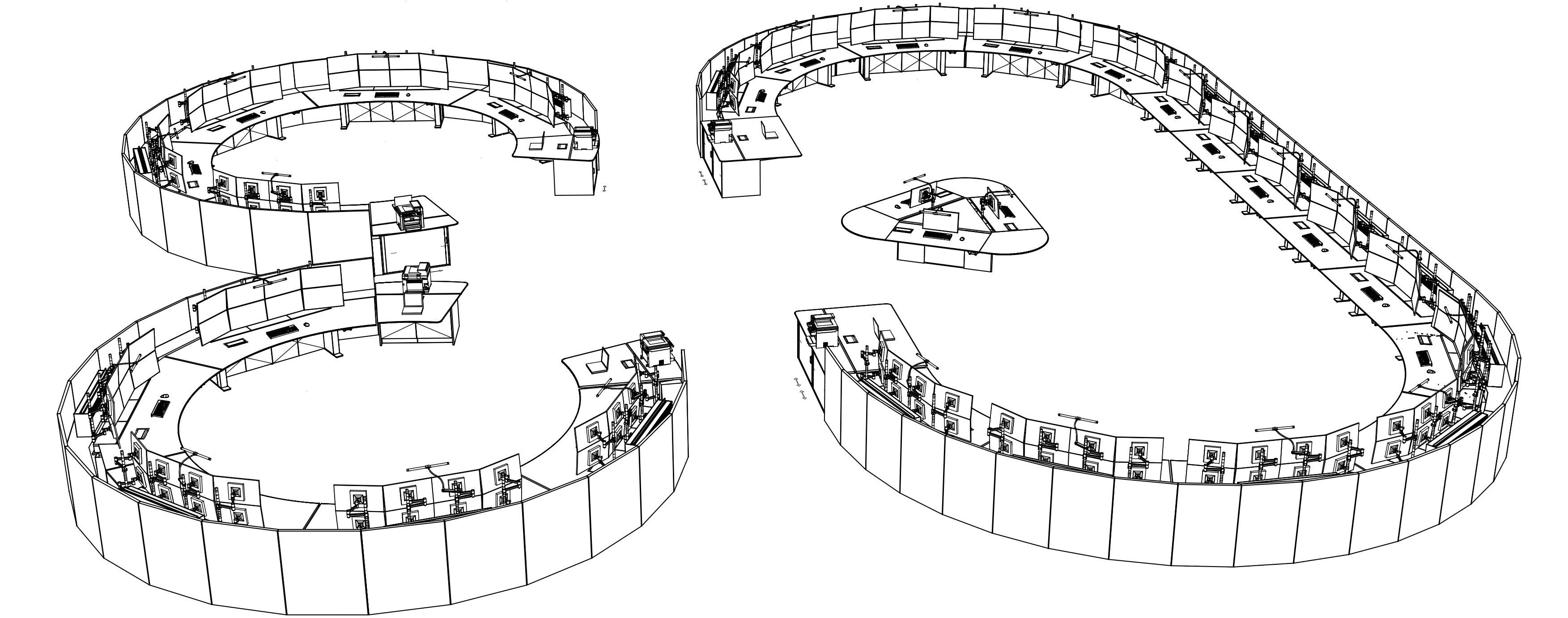}
   \caption{Console layout for the new FAIR control room.}
   \label{fig:paper_layout}
\end{figure}

A new control room for FAIR is under construction and expected to be operational in \SI{2026}{}~\cite{Vossberg:IPAC17-TUPIK047}. It will serve as the central control facility for all accelerator systems, encompassing UNILAC, SIS18, the GSI storage ring program and the FAIR complex. With an area of approx. \SI{600}{\square\meter}, it will be about \SI{2}{\times} larger than the current GSI control room and equipped with \SI{25}{} freely configurable workstations (Fig.~\ref{fig:paper_layout}).  

System control will be fully digital; no analog cabling is planned. 
Therefore, the entire injector chain is being migrated to the FAIR control system.
With the exception of UNILAC, this transition is largely complete~\cite{Krepp:ICALEPCS21-MOPV013, Ondreka:IPAC19-MOPTS035, litvinov:icalepcs2023-tupdp019, Herfurth:COOL17-THM13}.
UNILAC is currently being upgraded as part of the Injector Controls Upgrade Project~\cite{gerhard:icalepcs2023-tupdp018}.
A prototype test has already been successfully completed, with further tests scheduled for July~\SI{2025}{} and February~\SI{2026}{}.
These milestones will enable the \SI{2026}{} user beam time at GSI and FAIR hardware commissioning to be carried out from the new control room.

\section{Accelerator Commissioning}
\begin{figure}[h!]
\centering
\begin{adjustbox}{width=\columnwidth,trim=0.1cm 0 0 0,clip}
\begin{tikzpicture}

\foreach \x in {0.0,1.0,2.0,3.0} {
  \draw[dashed, gray!50, line width=0.3pt] (\x,-0.05) -- (\x,-1.9);
}
\foreach \x in {0.25,0.5,0.75,1.25,1.5,1.75,2.25,2.5,2.75,3.25,3.5,3.75} {
  \draw[dashed, gray!30, line width=0.2pt] (\x,-0.05) -- (\x,-1.9);
}

\foreach \x in {0.0,1.0,2.0,3.0} {
  \draw[line width=0.3pt] (\x,-1.9) -- (\x,-1.97);
}

\foreach \i/\name in {0/FCC, 1/HEBT, 2/SFRS, 3/SIS100, 4/Cryo}
    \node[anchor=east] at (0,-0.35-\i*0.35 + 0.07) {\scalebox{0.8}{\scriptsize \name}};

\draw[fill=tanngruen] (1.08,-0.354) rectangle (1.75,-0.234);

\draw[fill=tanngruen] (1.0,-0.704) rectangle (2.58,-0.584);
\draw[fill=red!80]   (2.58,-0.704) rectangle (3.0,-0.584);

\draw[fill=tanngruen] (1.67,-1.054) rectangle (2.67,-0.934);
\draw[fill=red!80]    (2.67,-1.054) rectangle (3.0,-0.934);
\draw[fill=blue!60]   (2.25,-1.110) rectangle (2.58,-0.994);

\draw[fill=tanngruen] (1.42,-1.404) rectangle (3.5,-1.284);
\draw[fill=red!80]    (3.5,-1.404) rectangle (4.0,-1.284);
\draw[fill=blue!60]   (3.0,-1.460) rectangle (3.42,-1.340);

\draw[fill=blue!60] (0.5,-1.72) rectangle (2.33,-1.6);

\draw[->] (0,-1.97) -- (4.0,-1.97);
\foreach \x/\year in {0.5/2025, 1.5/2026, 2.5/2027, 3.5/2028}
    \draw (\x,-2.02) node[below] {\scalebox{0.8}{\scriptsize \year}};

\end{tikzpicture}
\end{adjustbox}

\vspace{0.15em}
\begin{flushleft}
\begin{tikzpicture}[baseline]
  \matrix[column sep=6pt, row sep=1pt] {
    \node[draw=none, fill=tanngruen, minimum width=0.3cm, minimum height=0.2cm] {}; &
    \node[anchor=west] {\scriptsize Hardware Commissioning (Phases 1–3)}; \\

    \node[draw=none, fill=red!80, minimum width=0.3cm, minimum height=0.2cm] {}; &
    \node[anchor=west] {\scriptsize Beam Commissioning}; \\

    \node[draw=none, fill=blue!60, minimum width=0.3cm, minimum height=0.2cm] {}; &
    \node[anchor=west] {\scriptsize Commissioning of Cryogenic System / Machine Cooldown}; \\
  };
\end{tikzpicture}
\end{flushleft}

\caption{FAIR Commissioning Schedule (2025--2028).}
\label{fig:fair-schedule}
\end{figure}
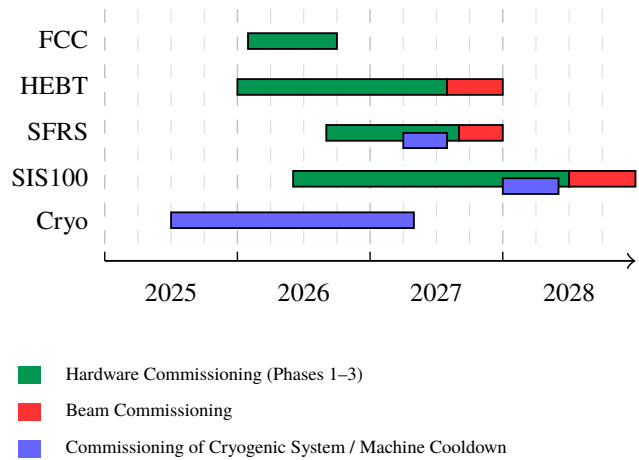

Commissioning preparations began in 2021 with the formation of the FAIR commissioning team. The process is structured into four phases:
\begin{enumerate}[itemsep=0pt]
    \item Local commissioning
    \item Remote \& system commissioning
    \item Integration tests and full dry-runs
    \item Beam commissioning
\end{enumerate}
For the first two phases, commonly referred to as device check-out, commissioning procedures for all system types were reviewed in detail. Each process step was documented together with its specific prerequisites and boundary conditions.
 Final acceptance tests before dry-runs will primarily use automated procedures executed by a software sequencer. These are under development and already tested at the existing GSI facility.
A high-level timeline of the commissioning activities is shown in Fig.~\ref{fig:fair-schedule}.

The pilot-beam commissioning will begin immediately after the final acceptance dry-run. Its goal is to deliver an ion beam defined by the Project Completion Parameters (PCPs) to a target or experiment. These parameters have been selected to ensure reliable achievement using the existing GSI injector chain while enabling the first physics experiment at NUSTAR or CBM. For this initial experiment, SIS100 will be operated with a single bunch instead of four. The PCPs for the respective beamlines are listed in Table~\ref{tab:pcp}. Completion of the first experiment marks the formal end of commissioning for the corresponding accelerator chain. Beam intensity will then be ramped up during subsequent operational phases.

\begin{table}[h!]
\caption{Project Completion Parameters}
\centering
\renewcommand{\arraystretch}{1.3}
\resizebox{\columnwidth}{!}{%
\begin{tabular}{|l|l|l|l|}
\hline
\textbf{Chain}        & \textbf{Ion}  & \textbf{Energy}  & \textbf{Ion-Intensity}  \\ \hline
SIS18–SFRS–NUSTAR           & $^{238}$U$^{73+}$ & 1~GeV/u          & $2 \times 10^{9}$ ion/s     \\ \hline
SIS100–SFRS–NUSTAR    & $^{238}$U$^{28+}$ & 1.4~GeV/u        & $2.5 \times 10^{10}$ ion/s   \\ \hline
SIS100–CBM            & $^{197}$Au$^{79+}$ & 11~GeV/u         & $1 \times 10^{7}$ ion/s     \\ \hline
\end{tabular}%
}
\label{tab:pcp}
\end{table}

\subsection{High Energy Beam Transport (HEBT)}

The HEBT system~\cite{Hagenbuck:IPAC15-THPF012} includes, in the Early Science stage, the beam line from SIS18 to the Super-FRS. It spans approximately \SI{300}{m}, with 82 magnets, 54 diagnostic devices, three safety beam plugs, and vacuum components such as pumps, valves, and gauges.

Initial hardware commissioning will begin in Q4~2025, once installations in building and in the tunnel are complete and infrastructure is operational. As work progresses in tunnel, further commissioning will continue section by section throughout 2026.
Remote/system commissioning and integration tests will follow, concluding with dry-runs in Q2~2027. In parallel, installation of the \SI{400}{m} HEBT First Science section will be finalized by mid-2027.
Beam commissioning is planned using fast-extracted SIS18 beam at \SIrange{12}{18}{Tm}, e.g., $^{40}\mathrm{Ar}^{18+}$ at \SI{1}{GeV/u}, with pilot intensities of $10^7$–$10^8$ ions per cycle.

Commissioning steps include measurement of beam parameters, optical tuning, element-by-element steering, and functional verification of beamline components. Once beam dumping on the Super-FRS catcher is confirmed, further tests such as kick response measurements, aperture scans, and dispersion measurements will be conducted. The beam can then be optimized for Super-FRS commissioning.

\subsection{Superconducting Fragment Separator (SFRS)}

\begin{figure}[!htb]
   \centering
   \includegraphics[width=1.0\columnwidth]{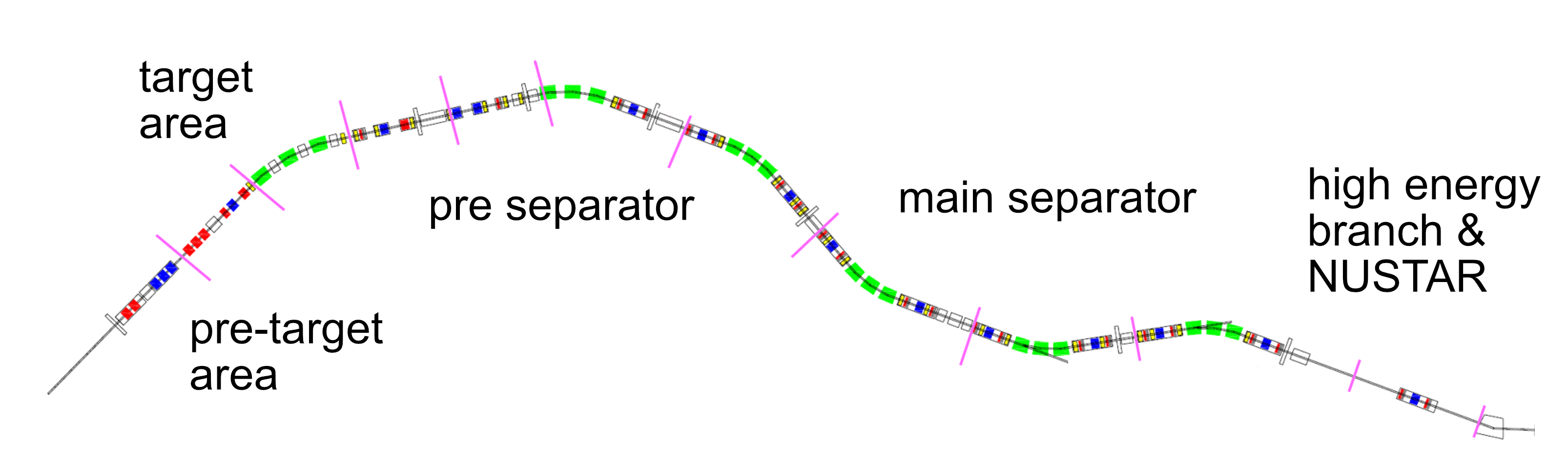}
   \caption{Super-FRS in its early science configuration.}
   \label{fig:sfrs_overview}
\end{figure}

The Super-FRS, in its Early Science configuration consisting only of the main branch (see Fig.~\ref{fig:sfrs_overview}), is expected to complete commissioning by the end of \SI{2027}{} in preparation for the first the NUSTAR experiment. This configuration comprises 150 racks, 30{,}000 cables, more than 120 magnetic elements, 40 drives, and 60 beam diagnostic devices installed across 25 vacuum sections. 

The superconducting magnets~\cite{sugita:ipac2023-wepm068} are cooled via eight cryogenic branches and 45 feed boxes.
The initial phases of commissioning will run in parallel with ongoing installation during \SI{2025}{} and \SI{2026}{}. All cryogenic sections are planned to be cooled down in \SI{2027}{}, enabling system dry-runs in Q3 and beam commissioning in Q4. The objectives of beam commissioning are to characterize the ion-optical parameters, validate the fragment selection and identification capabilities, and ensure all systems are fully functional and safe for delivering reliable fragment beams for the \SI{2028}{} Early Science campaign.

\subsection{The Heavy-Ion Synchrotron SIS100}

SIS100 is a heavy-ion synchrotron equipped with fast-ramped superconducting magnets. Its goal is to provide full cycling flexibility, similar to a room-temperature synchrotron~\cite{spiller:ipac2023-mopa062}. However, fast ramping induces dynamic losses in the cold mass, especially in the yokes of the superferric magnets, resulting in significant variations in cryogenic heat load depending on the operational cycle.

Operation modes vary by user requirements: either a long extraction flat-top for slow extraction and low AC loss, or a triangular cycle with fast extraction and high AC loss~\cite{Petzenhauser:IPAC21-WEPAB348}. SIS100’s design addresses this through two-phase \SI{4.5}{\kelvin} liquid helium cooling in the supply header and maximized helium gas return for efficient cryoplant operation. Hydraulic circuits in the quadrupole systems are balanced via calibrated flow restrictors, while LHe pumps and phase separators recover liquid helium from the return line.
A separate hydraulic circuit independently cools the ultra-high vacuum (UHV) system. This complex thermal management requires detailed cold commissioning tailored to different use cases. In the long term, predictive heat-load modeling and proactive control of hydraulic circuits via the accelerator control system are planned.

Cold commissioning is scheduled to begin in early \SI{2028}{} after cryomagnet installation. Initial steps will include commissioning of the main power converters and quench protection systems at various current levels, followed by user-specific operation cycles.

Conventional systems in the room-temperature straights will be commissioned in parallel with other FAIR components. RF systems are currently undergoing offline testing. Specialized devices such as the ramped bipolar extraction kicker will be fully integrated and FAT-tested at the manufacturer’s site.

\section{NUSTAR and CBM - Experiments}

The NUSTAR experimental collaboration, one of the four core scientific pillars at FAIR, focuses on the study of NUclear STructure, Astrophysics, and Reactions~\cite{KalantarNayestanaki2018}. Comprising more than 600 members from more than 140 international institutes, NUSTAR includes several sub-collaborations using a wide range of state-of-the-art instruments to explore different aspects of the overarching science program through complementary experimental methods.

Since 2019, NUSTAR experiments have been successfully conducted at existing GSI facilities as part of the FAIR Phase-0 program, during which many components were brought into operation. Three sub-collaborations, R3B, HISPEC/DESPEC, and the Super-FRS Experiment Collaboration, will be the first to relocate to the FAIR site for Early Science. These groups will perform new experiments using exotic secondary beams provided by the Super-FRS.

Installation of NUSTAR components in the High-Energy Cave will begin in 2026, followed by offline commissioning, integrated system tests, and full in-beam commissioning with Super-FRS beams. The NUSTAR Early Science physics program will commence once all systems are verified to be fully operational.

The Compressed Baryonic Matter (CBM) experiment aims to explore the phase structure of strong interaction (QCD) matter at large net-baryon densities and moderate temperatures using heavy-ion collisions in the energy range \(\sqrt{s_{\mathrm{NN}}} = 2.9\,\text{–}\,4.9\,\mathrm{GeV}\). CBM is a fixed-target experiment,
equipped with fast and radiation-hard detector systems and an advanced triggerless data acquisition scheme. The CBM will collect data at interaction rates of up to 10 MHz by performing online reconstruction and event selection, thus allowing measurements of rare probes not accessible so far in this energy range. These include: multi-strange hadron production and their flow, higher-order cumulants, dileptons, as well as double-strange hypernuclei. The installation of the experimental infrastructure commenced in 2023 and will proceed through 2027, culminating in the installation of the CBM detectors and the start of global commissioning. The SIS100 synchrotron and CBM experiment are expected to begin operations in late 2028.

\section{CONCLUSION}

The FAIR project is entering a crucial phase with the transition from installation to commissioning. A comprehensive strategy has been developed to ensure structured, stepwise commissioning of all systems, beginning with the cryogenic infrastructure and progressing through local hardware commissioning, integration tests, and finally beam commissioning. 
Key infrastructure such as the FAIR Control Centre and cryogenic plant will be operational by 2026. The outlined timeline provides a realistic path towards first science while maintaining parallel operation of GSI systems. 

\section{ACKNOWLEDGEMENTS}

The authors would like to thank Jörg Wenninger, Ciprian Plostinar, and Matthias Scholz for their continuous support and valuable advice.  
Preparatory conceptual work by the FAIR Commissioning \& Controls Working Group~\cite{Steinhagen:HB2018-TUA2WD01} is also acknowledged.

\ifboolexpr{bool{jacowbiblatex}}%
	{\printbibliography}%
	{
	
} 
\end{document}